\documentclass[aps]{revtex4}
\usepackage{graphicx}
\usepackage{graphicx}

\newcommand*{\be}
{\begin{equation}}
\newcommand*{\ee}
{\end{equation}}

\begin{document}

\title{Critical temperature for first-order phase transitions in confined systems}
\author{C.A. Linhares$^{a}$, A. P. C. Malbouisson$^{b}$,
Y.W. Milla$^{c}$, I. Roditi$^{b}$}
\affiliation{$^{a}$Instituto de F\'{\i}sica, Universidade do Estado do Rio de Janeiro, 
Rua S\~ao Francisco Xavier, 524, 20559-900 Rio de Janeiro,
RJ,Brazil\\
$^{b}$Centro Brasileiro de Pesquisas Fí sicas, Rua Dr.
Xavier Sigaud 150, 
22290-180 Rio de Janeiro, RJ, Brazil\\
$^{c}$Instituto de F\'{\i}sica Teorica-IFT/UNESP 
Rua Pamplona 145, 01405-900, S\~ao Paulo, SP, Brazil}

\begin{abstract}
\noindent We consider the Euclidean $D$-dimensional $-\lambda |\varphi
|^4+\eta |\varphi |^6$ ($\lambda ,\eta >0 $) model with $d$ ($d\leq D$) compactified dimensions.
Introducing temperature by means of the Ginzburg--Landau prescription in the
mass term of the Hamiltonian, this model can be interpreted as describing a
first-order phase transition for a system in a region of the $D$-dimensional
space, limited by $d$ pairs of parallel planes, orthogonal to the
coordinates axis $x_1,\,x_2,\,\ldots ,\,x_d$. The planes in each pair are
separated by distances $L_1,\;L_2,\;\ldots ,\,L_d$. We obtain an expression
for the transition temperature as a function of the size of the system, $%
T_c(\{L_i\})$, $i=1,\,2,\,\ldots ,d$. For $D=3$ we particularize this
formula, taking $L_1=L_2=\cdots =L_d=L$ for the physically interesting cases 
$d=1$ (a film), $d=2$ (an infinitely long wire having a square 
cross-section), and for $d=3$ (a cube). For completeness, the corresponding
formulas for second-order transitions are also presented. Comparison with
experimental data for superconducting films and wires shows  qualitative
agreement with our theoretical expressions.\\

\noindent PACS number(s): 03.70.+k, 11.10.-z
\end{abstract}

\maketitle

\section{Introduction}

Studies on field theory applied to second-order phase transitions have been 
done in the literature for a long time. A thorough account on the subject
can be found in Refs.~\cite
{cardy,affleck,lawrie1,lawrie2,brezin,radzi,calan,zinn}. A recent
application of similar ideas to bounded systems can also be found in Ref.~%
\cite{malbouisson5}. Under the assumption that information about general
features of the behavior of systems undergoing phase transitions can be
obtained in the approximation which neglects gauge field contributions in
the Ginzburg--Landau model, investigations have been done with an approach
different from the renormalization-group analysis. The system confined
between two parallel planes has been considered and using the formalism
developed in Refs. \cite{malbouisson2,malbouisson3,urucubaca} the way in
which the critical temperature for a second-order phase transition is
affected by the presence of confining boundaries has been investigated. In
particular, a study has been carried out on how the critical temperature of
a superconducting film depends on its thickness \cite
{luciano2,urucubaca,luciano}. Moreover, confined systems in regions of
three-dimensional space with some other shapes were also considered: grains
and wires \cite{malbouisson4,luciano3}. In all those cases a minimal size of
these regions can be determined for which the transition is still sustained.

In a previous article \cite{linhares} we have done a further step, by
considering in the simpler case of the system confined between two parallel
planes, the model which besides the quartic scalar field self-interaction, a
sextic one is also present. The model with both interactions taken together
leads to a renormalizable quantum field theory in three dimensions and, in
the context considered in Ref.~\cite{linhares}, it describes {\em first-order%
} phase transitions in films. In this paper we extend this formalism to a
general framework, considering the Euclidean $D$-dimensional $-\lambda
|\varphi |^4+\eta |\varphi |^6$ ($\lambda ,\eta >0$)model with $d$ ($d\leq D$) compactified
dimensions, from which we obtain general formulas for the dependence of the
transition temperature on the parameters delimiting the spatial region
within which the system is confined. Particularizing for $D=3$, we then consider
the superconducting material in the form of a film ($d=1$), of a wire ($d=2$), and
of a grain ($d=3$). We also present for comparison the corresponding
formulas for second-order transitions.

The usual Ginzburg--Landau Hamiltonian  considers only the term 
$\lambda \varphi ^4$ ($\lambda >0$). This model is known to lead to
second-order phase transitions. But a potential of the type $-\lambda \varphi
^4+\eta \varphi ^6$ ($\lambda $, $\eta >0$) \emph{ensures} that the system
undergoes a \emph{first-order} transition. See, for example, \cite{lebellac}.  
 In some of our previous papers (Refs. \cite{luciano,luciano2,urucubaca,luciano3,malbouisson4}) the $\lambda \varphi ^4$ model has
been used to determine a theoretical $T_c(L)\times L$ curve for films  wires and grains and a comparison  to experimental data for superconducting films has been done. In the
present work we wished to do the same with the extended model in order to
compare expected results from second- and first-order transitions. Of course, there are many other potentials that engender
first-order transitions, for instance, the Halperin--Lubensky--Ma potential \cite{HLM}, 
which induces first-order transitions in superconducting
materials by effect of integration over the gauge field and takes the form $%
-\alpha \varphi ^3+\beta \varphi ^4$. Also in ref. \cite{claude} it has been shown that the Halperin--Lubensky--Ma effect holds  for films, with a potential still different from both above. Our potential $-\lambda \varphi
^4+\eta \varphi ^6$ ($\lambda $, $\eta >0$) is just a simple choice to generate first-order transitions in the context of the Ginzburg-Landau theory. 

We consider, as in previous publications, that the system is a portion of
material of some size, the behavior of which in the critical region is to be
derived from a quantum field theory calculation of the dependence of the
physical mass parameter on its size. We start from the effective potential,
which is related to the physical mass through a renormalization condition.
This condition, however, reduces considerably the number of relevant Feynman
diagrams contributing to the mass, if one wishes to be restricted to
first-order terms in both coupling constants. In fact, just two diagrams
need to be considered in this approximation: a tadpole graph with the $\varphi
^4$ coupling (1 loop) and a ``shoestring'' graph with the $\varphi ^6$ coupling
(2 loops). No diagram with both couplings occurs. The size-dependence
appears from the treatment of the loop integrals. The dimensions of finite
extent are treated in momentum space using the formalism of Ref.~\cite
{malbouisson3}. 

It is worth to notice that for superconducting films with thickness $L$,
a qualitative agreement of our theoretical $L$-dependent critical temperature 
is found with experiments. This occurs in particular for thin films (in the case 
of first-order transitions) and for a wide range of values of $L$ for second-order
transitions \cite{linhares}. Moreover, the recently available experimental data
for superconducting wires \cite{shanenko,zgirski} are compatible with 
our theoretical prediction of the first-order critical temperature as a function of 
the transverse cross section of the wire.

The paper is organized as follows: In section II we present the model and
the general description of the $D$-dimensional Euclidean system with a
compactified $d$-dimensional subspace; for this, we make an adaptation of
the Matsubara formalism suited for our purposes. The contributions from the
relevant Feynman diagrams to the effective potential are then established.
Next, in section III, we exhibit expressions showing the size dependence of
the critical temperature for various shapes of confined materials.
Comparisons with experimental data for films and wires are shown. Finally,
in the the last section we present our conclusions.

\section{Effective potential with compactification of a $d$-dimensional
subspace}

We consider the scalar field model described by the Ginzburg--Landau
Hamiltonian density in a Euclidean $D$-dimensional space, including both $%
\varphi ^4$ and $\varphi ^6$ interactions, in the absence of external
fields, given by (in natural units, $\hbar =c=k_B=1$), 
\begin{equation}
{\cal H}=\frac 12\left| \partial _\mu \varphi \right| \left| \partial ^\mu
\varphi \right| +\frac 12m_0^2\left| \varphi \right| ^2-\frac \lambda 4%
\left| \varphi \right| ^4+\frac \eta 6\left| \varphi \right| ^6,
\label{lagrangiana}
\end{equation}
where $\lambda >0$ and $\eta >0$ are the {\em renormalized }quartic and
sextic self-coupling constants. Near criticality, the bare mass is given by $%
m_0^2=\alpha (T/T_0-1$), with $\alpha >0$ and $T_0$ being a parameter with
the dimension of temperature. Recall that the critical temperature for a
first-order transition described by the Hamiltonian above is higher than $%
T_0 $ \cite{lebellac}. This will be explicitly stated in Eq.~(\ref{crittemp}%
) below. Our purpose will be to develop the general case of compactifying a $%
d$-dimensional subspace.

We thus consider the system in $D$ dimensions confined to a region of space
delimited by $d\leq D$ pairs of parallel planes. Each plane of a pair $j$ is
at a distance $L_j$ from the other member of the pair, $j=1,2,\ldots ,d$,
and is orthogonal to all other planes belonging to distinct pairs $\{i\}$, $%
i\neq j$. This may be pictured as a parallelepipedal box embedded in the $D$%
-dimensional space, whose parallel faces are separated by distances $%
L_1,L_2,\ldots ,L_d$. We use Cartesian coordinates ${\bf r}=(x_1,...,x_d,%
{\bf z})$, where ${\bf z}$ is a $(D-d)$-dimensional vector, with
corresponding momentum ${\bf k}=(k_1,...,k_d,{\bf q})$, ${\bf q}$ being a $%
(D-d)$-dimensional vector in momentum space. The generating functional of
Schwinger functions is written in the form 
\begin{equation}
Z=\int {\cal D\varphi \,D\varphi }^{*}\exp \left( -\int_0^{L_1}dx_1\cdots
\int_0^{L_d}dx_d\int d^{D-d}x\,{\cal H}(\left| \varphi \right| ,\left|
\nabla \varphi \right| )\right) ,
\end{equation}
with the field $\varphi (x_1,...,x_d,{\bf z})$ satisfying the condition of
confinement inside the box, $\varphi (x_i\leq 0,{\bf z})=\varphi (x_i\geq 0,%
{\bf z})=$ const. Then, following the procedure developed in Ref.~\cite
{malbouisson3}, we are allowed to introduce a generalized Matsubara
prescription, performing the following multiple replacements
(compactification of a $d$-dimensional subspace), 
\begin{equation}
\int \frac{dk_i}{2\pi }\rightarrow \frac 1{L_i}\sum_{n_i=-\infty }^{+\infty
}\;;\;\;\;\;\;\;k_i\rightarrow \frac{2n_i\pi }{L_i}\;,\;\;i=1,2...,d.
\label{Matsubara1}
\end{equation}
Notice that compactification can be implemented in different ways, as for
instance by imposing specific conditions on the fields at spatial
boundaries. We here choose periodic boundary conditions.

We emphasize, however, that we are considering a Euclidean field theory in $%
D $ {\em purely} spatial dimensions. Therefore, we are {\em not} working
within the framework of finite-temperature field theory. Here, the
temperature is introduced in the mass term of the Hamiltonian by means of
the usual Ginzburg--Landau prescription.

In principle, the effective potential for systems with spontaneous symmetry
breaking is obtained, following the analysis introduced in Ref.~\cite
{coleman} (a new approach to this theorem is presented in Ref.~\cite
{malbouisson}), as an expansion in the number of loops in Feynman diagrams.
Accordingly, to the free propagator and to the no-loop (tree) diagrams for
both couplings, radiative corrections are added, with increasing number of
loops. Thus, at the 1-loop approximation, we get the infinite series of
1-loop diagrams with all numbers of insertions of the $\varphi ^4$ vertex
(two external legs in each vertex), plus the infinite series of 1-loop
diagrams with all numbers of insertions of the $\varphi ^6$ vertex (four
external legs in each vertex), plus the infinite series of 1-loop diagrams
with all kinds of mixed numbers of insertions of $\varphi ^4$ and $\varphi
^6 $ vertices. Analogously, we should include all those types of insertions
in diagrams with 2 loops, etc. However, instead of undertaking this
computation, in our approximation we restrict ourselves to the lowest terms
in the loop expansion. We recall that the gap equation we are seeking is
given by the renormalization condition in which the physical squared mass is
defined as the second derivative of the effective potential $U(\varphi _0)$
with respect to the classical field $\varphi _0$, taken at zero field, 
\begin{equation}
\left. \frac{\partial ^2U(\varphi _0)}{\partial |\varphi _0|{}^2}\right|
_{\varphi _0=0}=m^2.  \label{renorm}
\end{equation}
Within our approximation, we do not need to take into account the
renormalization conditions for the interaction coupling constants, i.e.,
they may be considered as already renormalized when they are written in the
Hamiltonian.

At the 1-loop approximation, the contribution of loops with only $%
|\varphi_0| ^4$ vertices to the effective potential is obtained directly
from \cite{malbouisson3}, as an adaptation of the Coleman--Weinberg
expression after compactification in $d$ dimensions. In this case, we start
from the well-known expression for the one-loop contribution to the
zero-temperature effective potential in unbounded space \cite{malbouisson3}, 
\begin{equation}
U_1(\varphi _0)=\sum_{s=1}^\infty \frac{(-1)^{s+1}}{2s}\left[ \frac{\lambda
|\varphi _0|^2}{2}\right] ^s\int \frac{d^Dk}{(k^2+m^2)^s},  \label{potefet0}
\end{equation}
where $m$ is the {\em physical} mass.

In the following, to deal with dimensionless quantities in the
regularization procedures, we introduce parameters $c^2=m^2/4\pi ^2\mu
^2,\;\;(L_i\mu )^2=a_i^{-1},\;\;g_1=(-\lambda /16\pi ^2\mu
^{4-D}),\;\;|\varphi _0/\mu ^{D-2}|^2=|\varphi _0|^2$, where $\varphi _0$ is
the normalized vacuum expectation value of the field (the classical field)
and $\mu $ is a mass scale. In terms of these parameters and performing the
Matsubara replacements (\ref{Matsubara1}), the one-loop contribution to the
effective potential can be written in the form 
\begin{eqnarray}
U_1(\varphi _0,a_1,...,a_d) &=&\mu ^D\sqrt{a_1\cdots a_d}\,\sum_{s=1}^\infty 
\frac{(-1)^{s+1}}{2s}g_1^s|\varphi _0|^{2s}  \nonumber  \label{potefet1} \\
&&\times \sum_{n_1,...,n_d=-\infty }^{+\infty }\int \frac{d^{D-d}q}{%
(a_1n_1^2+\cdots +a_dn_d^2+c^2+{\bf q}^2)^s}.  \label{potefet1}
\end{eqnarray}
The parameter $s$ counts the number of vertices on the loop.

It is easily seen that only the $s=1$ term contributes to the
renormalization condition (\ref{renorm}). It corresponds to the tadpole
diagram. It is then also clear that all $|\varphi _0|^6$-vertex and mixed $%
|\varphi _0|^4$- and $|\varphi _0|^6$-vertex insertions on the 1-loop diagrams do
not contribute when one computes the second derivative of similar
expressions with respect to the field at zero field: only diagrams with two
external legs should survive. This is impossible for a $|\varphi _0|^6$-vertex
insertion at the 1-loop approximation. Therefore, the first contribution
from the $|\varphi _0|^6$ coupling must come from a higher-order term in the
loop expansion. Two-loop diagrams with two external legs and only $|\varphi
_0|^4$ vertices are of second order in its coupling constant, and we neglect
them, as well as all possible diagrams with vertices of mixed type. However,
the 2-loop shoestring diagram, with only one $|\varphi _0|^6$ vertex and two
external legs is a first-order (in $\eta $) contribution to the effective
potential, according to our approximation.

In short, we consider the physical mass as defined at first-order in both
coupling constants, by the contributions of radiative corrections from only
two diagrams: the tadpole and the shoestring diagrams.

The tadpole contribution reads (putting $s=1$ in Eq.~(\ref{potefet1})) 
\begin{eqnarray}
U_1(\varphi _0,a_1,...,a_d) &=&\mu ^D\sqrt{a_1\cdots a_d}\,\frac 12\,g_1|\varphi
_0|^2  \nonumber  \label{potefet1} \\
&&\times \sum_{n_1,...,n_d=-\infty }^{+\infty }\int \frac{d^{D-d}q}{{\bf q}%
^2+a_1n_1^2+\cdots +a_dn_d^2+c^2}.  \label{potf1n}
\end{eqnarray}
The integral over the $D-d$ noncompactified momentum variables is performed
using the well-known dimensional regularization formula \cite{zinn} 
\begin{equation}
\int \frac{d^lq}{{\bf q}^2+M}=\frac{\Gamma (1-\frac l2)\pi ^{l/2}}{M^{1-l/2}}%
;  \label{dimreg}
\end{equation}
for $l=D-d$, we obtain 
\begin{equation}
U_1(\varphi _0,a_1,...,a_d)=\mu ^D\sqrt{a_1\cdots a_d}\,\sum_{s=1}^\infty
f(D,d)g_1|\varphi _0|^2Z_d^{c^2}\left( \frac{2-D+d}2;a_1,...,a_d\right) ,
\label{potefet2}
\end{equation}
where 
\begin{equation}
f(D,d)=\pi ^{(D-d)/2}\frac 12\Gamma \left( 1-\frac{D-d}2\right)
\end{equation}
and $Z_d^{c^2}(\nu ;a_1,...,a_d)$ are Epstein--Hurwitz zeta-functions, valid
for Re$(\nu )>d/2$, defined by 
\begin{eqnarray}
Z_d^{c^2}(\nu ;a_1,...,a_d) &=&\sum_{n_1,...,n_d=-\infty }^{+\infty
}(a_1n_1^2+\cdots +a_dn_d^2+c^2)^{-\nu }  \nonumber  \label{zeta} \\
&=&\frac 1{c^{2\nu }}+2\sum_{i=1}^d\sum_{n_i=1}^\infty (a_in_i^2+c^2)^{-\nu
}+2^2\sum_{i<j=1}^d\sum_{n_i,n_j=1}^\infty (a_in_i^2+a_jn_j^2+c^2)^{-\nu
}+\cdots  \nonumber \\
&&+2^d\sum_{n_1,...,n_d=1}^\infty (a_1n_1^2+\cdots +a_dn_d^2+c^2)^{-\nu }.
\end{eqnarray}
Next, we can proceed to generalizing to several dimensions the mode-sum
regularization prescription described in Ref.~\cite{elizalde}. This has been
done in Ref.~\cite{malbouisson3} and it results that the multidimensional
Epstein--Hurwitz function has an analytic extension to the whole $\nu $
complex plane, which may be written as (remembering that $a_i=(L_i\mu )^{-2}$%
) 
\begin{eqnarray}
Z_d^{c^2}(\nu ;a_1,...,a_d) &=&\frac{2^{\nu -\frac d2+1}\pi ^{2\nu -\frac d2}%
}{\sqrt{a_1\cdots a_d}\,\Gamma (\nu )}\left[ 2^{\nu -\frac d2-1}\left( \frac 
m\mu \right) ^{d-2\nu }\Gamma \left( \nu -\frac d2\right) \right.  \nonumber
\label{zeta4} \\
&&\left. +2\sum_{i=1}^d\sum_{n_i=1}^\infty \left( \frac m{\mu ^2L_in_i}%
\right) ^{\frac d2-\nu }K_{\nu -\frac d2}(mL_in_i)+\cdots \right.  \nonumber
\\
&&\left. +2^d\sum_{n_1,...,n_d=1}^\infty \left( \frac m{\mu ^2\sqrt{%
L_1^2n_1^2+\cdots +L_d^2n_d^2}}\right) ^{\frac d2-\nu }K_{\nu -\frac d2%
}\left( m\sqrt{L_1^2n_1^2+\cdots +L_d^2n_d^2}\right) \right] ,  \nonumber \\
&&  \label{zeta4}
\end{eqnarray}
where the $K_\nu $ are Bessel functions of the third kind. Taking $\nu
=(2-D+d)/2$ in Eq.~(\ref{zeta4}), we obtain from Eq.~(\ref{potefet2}) the
tadpole part of the effective potential in $D$ dimensions with a
compactified $d$-dimensional subspace: 
\begin{eqnarray}
U_1(\varphi _0,L_1,...,L_d) &=&\frac{\lambda |\varphi _0|^2}{2\,\left( 2\pi
\right) ^{D/2}}\left[ 2^{-D/2-1}m^{D-2}\Gamma \left( \frac{2-D}2\right)
+\sum_{i=1}^d\sum_{n_i=1}^\infty \left( \frac m{L_in_i}\right)
^{D/2-1}K_{D/2-1}(mL_in_i)\right.  \nonumber \\
&&\left. +2\sum_{j<i=1}^d\sum_{n_i,n_j=1}^\infty \left( \frac m{\sqrt{%
L_i^2+L_j^2}}\right) ^{D/2-1}K_{D/2-1}\left( m\sqrt{L_i^2+L_j^2}\right)
+\cdots \right.  \nonumber \\
&&\left. +2^{d-1}\sum_{n_1,\ldots ,n_d=1}^\infty \left( \frac m{\sqrt{%
L_1^2n_1^2+\cdots +L_d^2n_d^2}}\right) ^{D/2-1}K_{D/2-1}\left( m\sqrt{%
L_1^2n_1^2+\cdots +L_d^2n_d^2}\right) \right] ,  \nonumber \\
&&  \label{potefet3}
\end{eqnarray}
where we have returned to the original variables, $\varphi _0$, $\lambda $,
and $L_i$.

We now turn to the 2-loop shoestring diagram contribution to the effective
potential, using again the Matsubara-modified Feynman rule prescription for
the compactified dimensions. In unbounded space ($L_i=\infty $), it reads 
\begin{equation}
U_2(\varphi _{0})=\frac{\eta |\varphi _0|^{2}}{16}\left[ \int \frac{d^Dq}{\left(
2\pi \right) ^D}\frac 1{q^2+m^2}\right] ^2,
\end{equation}
which, after the compactification of $d$ dimensions of linear extensions $%
L_i $, $i=1,\ldots ,d$, becomes 
\begin{eqnarray}
U_2(\varphi _0,a_1,\ldots ,a_d) &=&\frac 12g_2|\varphi _0|^{2}\mu ^{2D-2}a_1\cdots
a_d \pi ^{D-d}  \nonumber \\
&&\times \left[ \Gamma \left( \frac{2-D+d}2\right) \sum_{n_1,\ldots
,n_d=-\infty }^\infty \frac 1{(a_1n_1^2+\cdots +a_dn_d^2+c^2)^{(2-D+d)/2}}%
\right] ^2,
\end{eqnarray}
where we have defined $\varphi _0$ and $a_i$ as before and the dimensionless
quantity $g_2=(\eta /8\cdot 16\pi ^4\mu ^{6-2D})$. Eq.~(\ref{dimreg}) was
also used. The multiple sum above is again the Epstein--Hurwitz zeta
function, $Z_d^{c^2}\left(\frac{2-D+d}2;a_1\cdots a_d\right)$, given by Eq.~(%
\ref{zeta4}) for $\nu =(2-D+d)/2$. In terms of the original variables, $%
\varphi$, $\eta$, and $L_i$, we then have 
\begin{eqnarray}
U_2(\varphi _0,L_1,\ldots ,L_d) &=&\frac{\eta |\varphi _0|^2}{4\,\left( 2\pi
\right) ^D\,}\left[ 2^{-1-D/2}m^{D-2}\Gamma \left( \frac{2-D}2\right) \right.
\nonumber \\
&&+\sum_{i=1}^d\sum_{n_i=1}^\infty \left( \frac m{L_in_i}\right)
^{D/2-1}K_{D/2-1}(mL_in_i)+\cdots  \nonumber \\
&&\left. +2^{d-1}\sum_{n_1,...,n_d=1}^\infty \left( \frac m{\sqrt{%
L_1^2n_1^2+\cdots +L_d^2n_d^2}}\right) ^{D/2-1}K_{D/2-1}\left(m\sqrt{%
L_1^2n_1^2+\cdots +L_d^2n_d^2}\right)\right] ^2.  \label{potefet4}
\end{eqnarray}
Notice that in both Eqs.~(\ref{potefet3}) and (\ref{potefet4}) there is a
term proportional to $\Gamma \left( \frac{2-D}2\right) $ which is divergent
for even dimensions $D\geq 2$ and should be subtracted in order to obtain
finite physical parameters. For odd $D$, the above gamma function is finite,
but we also subtract its term (corresponding to a finite renormalization)
for the sake of uniformity. After subtraction we get 
\begin{eqnarray}
U_1^{{\rm (Ren)}}(\varphi _0,L_1,\ldots ,L_d) &=&\frac{\lambda |\varphi
_0|^{2}}{2\,\left( 2\pi \right) ^{D/2}}\left[
\sum_{i=1}^d\sum_{n_i=1}^\infty \left( \frac m{L_in_i}\right)
^{D/2-1}K_{D/2-1}(mL_in_i)\right.  \nonumber \\
&&+2\sum_{i<j=1}^d\sum_{n_i,n_j=1}^\infty \left( \frac m{\sqrt{%
L_i^2n_i^2+L_j^2n_j^2}}\right) ^{D/2-1}K_{D/2-1}\left(m\sqrt{%
L_i^2n_i^2+L_j^2n_j^2}\right)+\cdots  \nonumber \\
&&\left. +2^{d-1}\sum_{n_1,\ldots ,n_d=1}^\infty \left( \frac m{\sqrt{%
L_1^2n_1^2+\cdots +L_d^2n_d^2}}\right) ^{D/2-1}K_{D/2-1}\left(m\sqrt{%
L_1^2n_1^2+\cdots +L_d^2n_d^2}\right)\right]  \nonumber \\
&&  \label{u1ren}
\end{eqnarray}
and 
\begin{eqnarray}
U_2^{{\rm (Ren)}}(\varphi _0,L_1,\ldots ,L_d) &=&\frac{\eta |\varphi _0|^{2}%
}{4\,(2\pi )^D}\left[ \sum_{i=1}^d\sum_{n_i=1}^\infty \left( \frac m{L_in_i}%
\right) ^{D/2-1}K_{D/2-1}(mL_in_i)\right.  \nonumber \\
&&+2\sum_{i<j=1}^d\sum_{n_i,n_j=1}^\infty \left( \frac m{\sqrt{%
L_i^2n_i^2+L_j^2n_j^2}}\right) ^{D/2-1}K_{D/2-1}(m\sqrt{L_i^2n_i^2+L_j^2n_j^2%
})+\cdots  \nonumber \\
&&\left. +2^{d-1}\sum_{n_1,\ldots ,n_d=1}^\infty \left( \frac m{\sqrt{%
L_1^2n_1^2+\cdots +L_d^2n_d^2}}\right) ^{D/2-1}K_{D/2-1}(m\sqrt{%
L_1^2n_1^2+\cdots +L_d^2n_d^2})\right] ^2.  \nonumber \\
&&  \label{U2ren}
\end{eqnarray}

Then the physical mass with both contributions is obtained from Eq.~(\ref
{renorm}), using Eqs.~(\ref{u1ren}), (\ref{U2ren}) and also taking into
account the contribution at the tree level; it satisfies a generalized
Dyson--Schwinger equation depending on the extensions $L_i$ of the confining
box:

\begin{eqnarray}
m^2(\{L_i\}) &=&m_0^2-\frac \lambda {\left( 2\pi \right) ^{D/2}}\left[
\sum_{n_1=1}^\infty \left( \frac m{L_1n_1}\right)
^{D/2-1}K_{D/2-1}(mL_1n_1)+\cdots +\sum_{n_d=1}^\infty \left( \frac m{L_dn_d}%
\right) ^{D/2-1}K_{D/2-1}(mL_dn_d)\right.  \nonumber \\
&&+2\sum_{i<j=1}^d\sum_{n_i,n_j=1}^\infty \left( \frac m{\sqrt{%
L_i^2n_i^2+L_j^2n_j^2}}\right) ^{D/2-1}K_{D/2-1}\left( m\sqrt{%
L_i^2n_i^2+L_j^2n_j^2}\right) +\cdots  \nonumber \\
&&+\left. 2^{d-1}\sum_{n_1,\ldots ,n_d=1}^\infty \left( \frac m{\sqrt{%
L_1^2n_1^2+\cdots +L_d^2n_d^2}}\right) ^{D/2-1}K_{D/2-1}\left( m\sqrt{%
L_1^2n_1^2+\cdots +L_d^2n_d^2}\right) \right]  \nonumber \\
&&+\frac \eta {2(2\pi )^D}\left[ \sum_{n_1=1}^\infty \left( \frac m{L_1n_1}%
\right) ^{D/2-1}K_{D/2-1}(mL_1n_1)+\cdots +\sum_{n_d=1}^\infty \left( \frac m%
{L_dn_d}\right) ^{D/2-1}K_{D/2-1}(mL_dn_d)\right.  \nonumber \\
&&+2\sum_{i<j=1}^d\sum_{n_i,n_j=1}^\infty \left( \frac m{\sqrt{%
L_i^2n_i^2+L_j^2n_j^2}}\right) ^{D/2-1}K_{D/2-1}\left( m\sqrt{%
L_i^2n_i^2+L_j^2n_j^2}\right) +\cdots  \nonumber \\
&&\left. +2^{d-1}\sum_{n_1,\ldots ,n_d=1}^\infty \left( \frac m{\sqrt{%
L_1^2n_1^2+\cdots L_d^2n_d^2}}\right) ^{D/2-1}K_{D/2-1}\left( m\sqrt{%
L_1^2n_1^2+\cdots L_d^2n_d^2}\right) \right] ^2.  \nonumber \\
&&  \label{massren1}
\end{eqnarray}
A first-order transition occurs when all the three minima of the potential 
\begin{equation}
U(\varphi _0)=\frac 12m^2(\{L_i\})|\varphi _0|^2-\frac \lambda 4|\varphi
_0|^4+\frac \eta 6|\varphi _0|^6,  \label{potencial}
\end{equation}
where $m(\{L_i\})$ is the renormalized mass defined above, are
simultaneously on the line $U(\varphi _0)=0$. This gives the condition 
\begin{equation}
m^2(\{L_i\})=\frac{3\lambda ^2}{16\eta }.  \label{condicao}
\end{equation}
For $D=3$, the Bessel functions entering in the above equations have an
explicit form, $K_{1/2}(z)=\sqrt{\pi }e^{-z}/\sqrt{2z}$, which is to be
replaced in Eq.~(\ref{massren1}). Performing the resulting sums gives 
\begin{eqnarray}
m^2(\{L_i\}) &=&\alpha \left( \frac T{T_0}-1\right) +\frac \lambda {8\pi }%
\left[ \sum_{i=1}^d\frac 1{L_i}\ln \left( 1-e^{-m(L_i)L_i}\right)
-2\sum_{j<i=1}^d\sum_{n_i,n_j=1}^\infty \frac{e^{-m(L_i)\sqrt{%
L_i^2n_i^2+L_j^2n_j^2}}}{\sqrt{L_i^2n_i^2+L_j^2n_j^2}}-\cdots \right. \\
&&\left. -2^{d-1}\sum_{n_1,\ldots ,n_d=1}^\infty \frac{e^{-m(L_i)\sqrt{%
L_1^2n_1^2+\cdots +L_d^2n_d^2}}}{\sqrt{L_1^2n_1^2+\cdots +L_d^2n_d^2}}%
\right] +\frac{\eta \pi }{8\,(2\pi )^3}\left[ \sum_{i=1}^d\frac 1{L_i}\ln
\left( 1-e^{-m(L_i)L_i}\right) \right.  \nonumber \\
&&\left. -2\sum_{j<i=1}^d\sum_{n_i,n_j=1}^\infty \frac{e^{-m(L_i)\sqrt{%
L_i^2n_i^2+L_j^2n_j^2}}}{\sqrt{L_i^2n_i^2+L_j^2n_j^2}}+...-2^{d-1}\sum_{n_1,%
\ldots ,n_d=1}^\infty \frac{e^{-m(L_i)\sqrt{L_1^2n_1^2+\cdots +L_d^2n_d^2}}}{%
\sqrt{L_1^2n_1^2+\cdots +L_d^2n_d^2}}\right] ^2.  \nonumber \\
&&  \label{massren2}
\end{eqnarray}
Then introducing the value of the mass, Eq.~(\ref{condicao}), in Eq.~(\ref
{massren2}), one obtains the critical temperature 
\begin{eqnarray}
T_c(\{L_i\}) &=&T_c\left\{ 1-\left( 1+\frac{3\lambda ^2}{16\eta \alpha }%
\right) ^{-1}\left\{ \frac \lambda {8\pi \alpha }\left[ \sum_{i=1}^d\frac 1{%
L_i}\ln \left( 1-e^{-\sqrt{\frac{3\lambda ^2}{16\eta }}L_i}\right)
-2\sum_{j<i=1}^d\sum_{n_i,n_j=1}^\infty \frac{e^{-m(L_i)\sqrt{%
L_i^2n_i^2+L_j^2n_j^2}}}{\sqrt{L_i^2n_i^2+L_j^2n_j^2}}\right. \right. \right.
\nonumber \\
&&\left. -2^{d-1}\sum_{n_1,\ldots ,n_d=1}^\infty \frac{e^{-\sqrt{\frac{%
3\lambda ^2}{16\eta }}\sqrt{L_1^2n_1^2+\cdots +L_d^2n_d^2}}}{\sqrt{%
L_1^2n_1^2+\cdots +L_d^2n_d^2}}\right] +\frac \eta {64\pi ^2\alpha }\left[
\sum_{i=1}^d\frac 1{L_i}\ln \left( 1-e^{-\sqrt{\frac{3\lambda ^2}{16\eta }}%
L_i}\right) \right.  \nonumber \\
&&\left. \left. \left. -2\sum_{j<i=1}^d\sum_{n_i,n_j=1}^\infty \frac{%
e^{-m(L_i)\sqrt{L_i^2n_i^2+L_j^2n_j^2}}}{\sqrt{L_i^2n_i^2+L_j^2n_j^2}}%
+2^{d-1}\sum_{n_1,\ldots ,n_d=1}^\infty \frac{e^{-\sqrt{\frac{3\lambda ^2}{%
16\eta }}\sqrt{L_1^2n_1^2+\cdots +L_d^2n_d^2}}}{\sqrt{L_1^2n_1^2+\cdots
+L_d^2n_d^2}}\right] ^2\right\} \right\} ,  \label{crittemp}
\end{eqnarray}
where 
\begin{equation}
T_c=T_0\left( 1+\frac{3\lambda ^2}{16\eta \alpha }\right)  \label{tc}
\end{equation}
is the bulk ($L_i\rightarrow \infty $) critical temperature for the
first-order phase transition.

\section{The film, the wire and the grain}

Having developed the general case of a $d$-dimensional compactified
subspace, it is now easy to obtain the specific formulas for particular
values of $d$. If we choose $d=1$, the compactification of just one
dimension, let us say, along the $x_1$-axis, we are considering that the
system is confined between two planes, separated by a distance $L_1=L$.
Physically, this corresponds to a film of thickness $L$ and we have that the
transition occurs at the critical temperature $T_c^{\text{film}}(L)$ given by

\begin{equation}
T_c^{\text{film}}(L)=T_c\left\{ 1-\left( 1+\frac{3\lambda ^2}{16\eta\alpha }%
\right)^{-1}\left[ \frac \lambda {8\pi \alpha L}\ln \left(1-e^{-\sqrt{\frac{%
3\lambda ^2}{16\eta }}L}\right)+\frac \eta {64\pi ^2\alpha L^2}\left(\ln
(1-e^{-\sqrt{\frac{3\lambda ^2}{16\eta }}L})\right)^2\right] \right\}.
\label{tcfilm}
\end{equation}

Let us now take the case $d=2$, in which the system is confined
simultaneously between two parallel planes a distance $L_1$ apart from one
another normal to the $x_1$-axis and two other parallel planes, normal to
the $x_2$-axis separated by a distance $L_2$. That is, the material is
bounded within an infinite wire of rectangular cross section $L_1\times L_2$%
. To simplify matters, we take $L_1=L_2=L$ in Eq. (\ref{crittemp}) with $d=2$%
, and the critical temperature is written in terms of $L$ as

\begin{eqnarray}
T_c^{\text{wire}}(L) &=&T_c\left\{ 1-\left( 1+\frac{3\lambda ^2}{16\eta
\alpha }\right) ^{-1}\left[ \frac \lambda {4\pi \alpha L}\left[ 2\ln
\left(1-e^{-L\sqrt{\frac{3\lambda ^2}{16\eta }}}\right)-2\sum_{n_1,n_2=1}^%
\infty \frac{e^{-L\sqrt{\frac{3\lambda ^2}{16\eta}}\sqrt{n_1^2+n_2^2}}}{%
\sqrt{n_1^2+n_2^2}}\right] \right.\right.  \nonumber \\
&&\left.\left. +\frac \eta {32\pi ^2\alpha L^2}\left( 2\ln \left(1-e^{-L%
\sqrt{\frac{3\lambda ^2}{16\eta }}}\right)-2\sum_{n_1,n_2=1}^\infty \frac{%
e^{-L\sqrt{\frac{3\lambda ^2}{16\eta }}\sqrt{n_1^2+n_2^2}}}{\sqrt{n_1^2+n_2^2%
}}\right) ^2\right] \right\} .  \label{tcwire}
\end{eqnarray}

Finally, we may compactify all three dimensions, which leaves us with a
system in the form of a cubic ``grain'' of some material. The dependence of
the critical temperature on its linear dimension $L_1=L_2=L_3=L$, is given
by putting $d=3$ in Eq.~(\ref{crittemp}): 
\begin{eqnarray}
T_c^{\text{grain}}(L) &=&T_c\left\{ 1-\left( 1+\frac{3\lambda ^2}{16\eta
\alpha }\right) ^{-1}\left\{ \frac \lambda {4\pi \alpha L}\left[ 3\ln \left(
1-e^{-L\sqrt{\frac{3\lambda ^2}{16\eta }}}\right)
-2\sum_{j<i=1}^3\sum_{n_i,n_j=1}^\infty \frac{e^{-\sqrt{\frac{3\lambda ^2}{%
16\eta }}L\sqrt{n_i^2+n_j^2}}}{\sqrt{n_i^2+n_j^2}}\right. \right. \right. 
\nonumber \\
&&\left. \left. \left. -4\sum_{n_1,\ldots ,n_3=1}^\infty \frac{e^{-L\sqrt{%
\frac{3\lambda ^2}{16\eta }}\sqrt{n_1^2+n_2^2+n_3^2}}}{\sqrt{%
n_1^2+n_2^2+n_3^2}}\right] +\frac \eta {32\pi ^2\alpha L^2}\left[ 3\ln
\left( 1-e^{-L\sqrt{\frac{3\lambda ^2}{16\eta }}}\right) \right. \right.
\right.  \nonumber \\
&&\left. \left. \left. -2\sum_{j<i=1}^3\sum_{n_i,n_j=1}^\infty \frac{%
e^{-m(L)L\sqrt{n_i^2+n_j^2}}}{\sqrt{n_i^2+n_j^2}}-4\sum_{n_1,\ldots
,n_3=1}^\infty \frac{e^{-L\sqrt{\frac{3\lambda ^2}{16\eta }}\sqrt{%
n_1^2+n_2^2+n_3^2}}}{\sqrt{n_1^2+n_2^2+n_3^2}}\right] ^2\right\} \right\} .
\label{tcgrain}
\end{eqnarray}

A similar work has been done for a {\em second-order} transition in either
films, wires or grains, obtained by the same methods from the $\lambda \varphi
^4$ Ginzburg--Landau model \cite{luciano3}. In this case the $\{L_i\}$%
-dependent physical mass has a simpler expression, 
\begin{eqnarray}
m_{\text{2nd}}^2(\{L_i\}) &=&m_0^2-\frac \lambda {\left( 2\pi \right) ^{D/2}}%
\left[ \sum_{n_1=1}^\infty \left( \frac m{L_1n_1}\right)
^{D/2-1}K_{D/2-1}(mL_1n_1)+\cdots +\sum_{n_d=1}^\infty \left( \frac m{L_dn_d}%
\right) ^{D/2-1}K_{D/2-1}(mL_dn_d)\right.  \nonumber \\
&&+2\sum_{i<j=1}^d\sum_{n_i,n_j=1}^\infty \left( \frac m{\sqrt{%
L_i^2n_i^2+L_j^2n_j^2}}\right) ^{D/2-1}K_{D/2-1}\left( m\sqrt{%
L_i^2n_i^2+L_j^2n_j^2}\right) +\cdots  \nonumber \\
&&+\left. 2^{d-1}\sum_{n_1,\ldots ,n_d=1}^\infty \left( \frac m{\sqrt{%
L_1^2n_1^2+\cdots +L_d^2n_d^2}}\right) ^{D/2-1}K_{D/2-1}\left( m\sqrt{%
L_1^2n_1^2+\cdots +L_d^2n_d^2}\right) \right] ,  \label{massa2nd}
\end{eqnarray}
from which, taking all $L_i$'s equal to $L$ and going to the limit $m_{\text{%
2nd}}^2(\{L_i\})\rightarrow 0$, formulas for the transition temperature for
films, wires and grains can be obtained. All of them have the same
functional dependence on the linear dimension $L$. In all cases studied
there, it is found that the boundary-dependent critical temperature
decreases linearly with the inverse of the linear dimension $L$, 
\begin{equation}
T_c^{\text{2nd}}(L)=T_0-\frac{C_d\lambda }{\alpha L},  \label{seg}
\end{equation}
where $\alpha $ and $\lambda $ are the Ginzburg--Landau parameters, $T_0$ is
the bulk transition temperature and $C_d$ is a constant equal to $1.1024$, $%
1.6571$ and $2.6757$ for $d=1$ (film), $d=2$ (square section wire) and $d=3$
(cubic grain), respectively.

Comparing Eqs.~(\ref{tcfilm}), (\ref{tcwire}) and (\ref{tcgrain}) with Eq.~(%
\ref{seg}), we see that in all the cases (a film, a wire or a grain), there
is a sharp contrast between the simple inverse linear behavior of $T_c(L)$
for second-order transitions and the rather involved dependence on $L$ of
the critical temperature for first-order transitions. These two types of
behavior prompt us to try to clarify the subject further, by comparing the
theoretical curves with experimental data for superconducting materials.
However, as far as we know, no available data exist for superconducting
grains. We shall thus consider the situations of bounded systems in the form
of a film or of a wire. In so doing, we can explicitly compare the forms of
the $T_c(L)$ curves for both first- and second-order transitions, and also
exhibit the degree of agreement between our theoretical expressions for the
first-order critical temperature and some experimental results obtained from
superconducting films and wires.

To start, we mention the generalization of Gorkov's \cite
{gorkov,abrikosov,kleinert} microscopic derivation for the $\lambda \varphi
^4$ model in order to include the additional interaction term $\eta \varphi
^6$ in the free energy \cite{linhares}. The interest here is to determine
the phenomenological constant $\eta $ as a function of the microscopic
parameters of the material, in an analogous way as has been done for the
constant $\lambda $ in the $\lambda \varphi ^4$ model. This leads to \cite
{linhares},

\begin{equation}
\alpha =1\,\;\;\;\;\lambda \approx 111.08\,\left( \frac{T_0}{T_F}\right)
^2,\;\;\;\;\eta \approx 8390\,\left( \frac{T_0}{T_F}\right) ^4,\;\;\;\;m_0^2=%
\frac T{T_0}-1,  \label{parametros3}
\end{equation}
where $T_F$ is the Fermi temperature and $T_0$ is the temperature parameter
introduced in Eq.~(\ref{lagrangiana}), which can be obtained from the
first-order bulk critical temperature by means of Eq.~(\ref{tc}).

By replacing the above constants in Eq.~(\ref{tcfilm}), we get the critical
temperature as a function of the film thickness and in terms of tabulated
microscopic parameters for specific materials.

We remark that Gorkov's original derivation of the phenomenological
constants is valid only for perfect crystals, where the electron mean free
path $l$ is infinite. However, we know that in many superconductors the
attractive interaction between electrons (necessary for pairing) is brought
about indirectly by the interaction between the electrons and the vibrating
crystal lattice (phonons). The presence of impurities within the crystal lattice modifies the interaction between electrons and phonons, with the consequence of making the electron mean free path finite. In fact the dieter is the sample, shorter the mean free path becomes \cite{kleinert}. The Ginzburg--Landau phenomenological
constants $\lambda $ and $\eta $ and the coherence length are somehow
related to the interaction of the electron pairs with the crystal lattice
and the impurities. A way of taking these facts into account preserving the
form of the Ginzburg--Landau free energy is to modify the intrinsic
coherence length $\xi _0$ and the coupling constants. Accordingly \cite
{kleinert}, $\xi _0\rightarrow \rho ^{1/2}\xi _0$, $\lambda \rightarrow
2\rho ^{-3/2}\lambda $ and $\eta \rightarrow 4\rho ^{-3}\eta $, where $\rho
\approx 0.18R^{-1}$, with $R=\xi _0/l$, where $\xi _0=0.13(\hbar v_F/k_BT_0)$%
. Then, it can be shown that Eq. (\ref{tcfilm}) becomes \cite{linhares}, 
\begin{eqnarray}
T_c^{\text{film}}(L) &=&T_c\left\{ 1-\left( 1+\frac{3\lambda ^2}{16\eta }%
\right) ^{-1}\left[ \frac{2R\lambda }{0.18\cdot 8\pi }\frac{\xi _0}L\ln
\left( 1-e^{-\frac L{\xi _0}\sqrt{\frac{3\lambda ^2}{16\eta }\frac R{0.18}}%
}\right) \right. \right.  \nonumber  \label{Tcritical2} \\
&&\left. \left. +\frac{4R^2\eta }{0.18^2\cdot 32\pi ^2}\left( \frac{\xi _0}L%
\right) ^2\left( \ln \left( 1-e^{-\frac L{\xi _0}\sqrt{\frac{3\lambda ^2}{%
16\eta }\frac R{0.18}}}\right) \right) ^2\right] \right\} .  \nonumber \\
&&
\end{eqnarray}

Also for realistic samples other effects, such that of the substrate over
which the superconductor film is deposited, should be taken into account. In
the context of our model, however, we are not able to describe such effects
at a microscopic level. We therefore assume that they will be translated in
changes on the values of the coupling constants $\lambda $ and $\eta $. So,
we propose as an {\em Ansatz} the rescaling of the constants in the form $%
\lambda \rightarrow a\lambda $ and $\eta \rightarrow a^2\eta $. We may still
combine both parameters $R$ and $a$ as $r=aR$. Eq.~(\ref{Tcritical2}) is
then written as 
\begin{eqnarray}
T_c^{\text{film}}(L) &=&T_c\left\{ 1-\left( 1+\frac{3\lambda ^2}{16\eta }%
\right) ^{-1}\left[ \frac{2r\lambda }{0.18\cdot 8\pi }\frac{\xi _0}L\ln
\left( 1-e^{-\frac L{\xi _0}\sqrt{\frac{3\lambda ^2}{16\eta }\frac R{0.18}}%
}\right) \right. \right.  \nonumber \\
&&\left. \left. +\frac{4r^2\eta }{0.18^2\cdot 32\pi ^2}\left( \frac{\xi _0}L%
\right) ^2\left( \ln \left( 1-e^{-\frac L{\xi _0}\sqrt{\frac{3\lambda ^2}{%
16\eta }\frac R{0.18}}}\right) \right) ^2\right] \right\} .
\label{Tcritical3}
\end{eqnarray}

In Fig.~1 we plot Eq.~(\ref{Tcritical3}) to show the behavior of the
transition temperature as a function of the thickness for a film made from
aluminum. The values for Al of the Fermi temperature, the bulk critical
temperature and Fermi velocity are $T_F=13.53\times 10^4$ K, $T_c=1.2$ K,
and $v_F=2.02\times 10^6m/s$, respectively. 
\begin{figure}[t]
\includegraphics[{height=7.0cm,width=8.5cm,angle=360}]{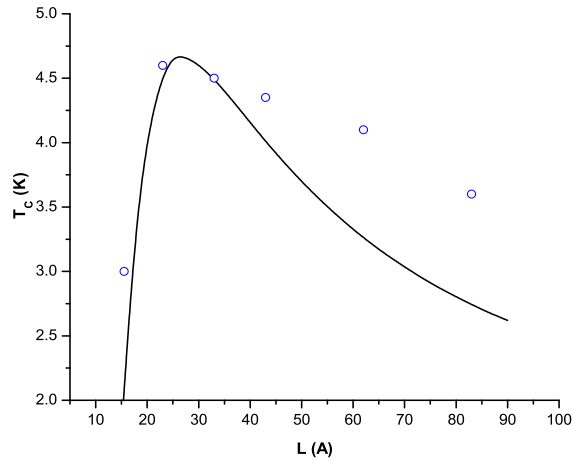}
\caption{Critical temperature $T_c^{\text{film}}$(K) as function of
thickness $L$ ($\AA $), from Eq.~(\ref{Tcritical3}) and data from Ref.~%
\protect\cite{strongin} for a superconducting film made from aluminum.}
\end{figure}

We see from the figure that the critical temperature grows from zero at a
nonnull minimal allowed film thickness above the bulk transition temperature 
$T_c$ as the thickness is enlarged, reaching a maximum and afterwards
starting to decrease, going asymptotically to $T_c$ as $L\rightarrow \infty $%
. We also plot for comparison some experimental data obtained from Ref.~\cite
{strongin}. We see that our theoretical curve is in qualitatively good
agreement with measurements, especially for thin films.
\begin{figure}[t]
\includegraphics[{height=7.0cm,width=8.5cm,angle=360}]{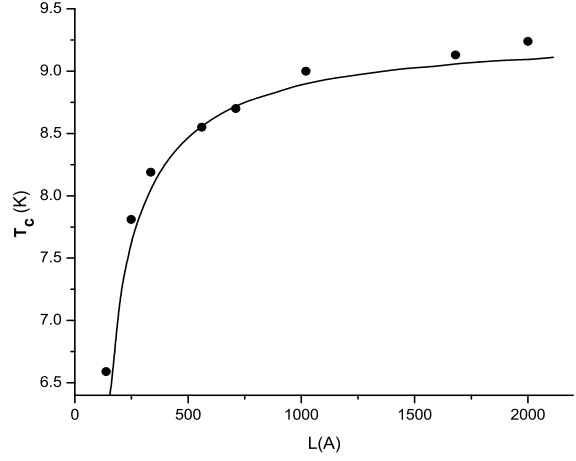}
\caption{Critical temperature $T_c$ (K) as a function of the thickness $L$ ($%
\AA $) for a second-order transition, as theoretically predicted
in Ref.~\protect\cite{luciano2}. Dots are experimental data taken
from Ref.~\protect\cite{itoh} for a superconducting film made from
niobium.}
\end{figure}

This behavior may be contrasted with the one shown by the critical
temperature for a {\em second-order} transition. In this case, the critical
temperature increases monotonically from zero, again corresponding to a
finite minimal film thickness, going asymptotically to the bulk transition
temperature as $L\rightarrow \infty $. This is illustrated in Fig.~2,
adapted from Ref.~\cite{luciano2}, with experimental data from \cite{itoh}.
(Such behavior has also been experimentally found by some other groups for a
variety of transition-metal materials, see Refs.~\cite
{raffy,minhaj,pogrebnyakov}.) Since in the present work a first-order
transition is explicitly assumed, it is tempting to infer that the
transition described in the experiments of Refs.~\cite{strongin,ramallo} is
first order. In other words, one could say that an experimentally observed
behavior of the critical temperature as a function of the film thickness may
serve as a possible criterion to decide about the order of the
superconductivity transition: a monotonically increasing critical
temperature as $L$ grows would indicate that the system undergoes a
second-order transition, whereas if the critical temperature presents a
maximum for a value of $L$ larger than the minimal allowed one, this would
be signalling the occurrence of a first-order transition.

Let us now consider a sample of superconducting material in the form of an
infinitely long wire with a cross section of side $L$. The same arguments
and rescaling procedures used precedingly for films apply equally in the
present situation. In this way, Eq. (\ref{tcwire}) is accordingly modified.
It assumes the form
\begin{eqnarray}
T_c^{\text{wire}}(L) &=&T_c\left\{ 1-\left( 1+\frac{3\lambda ^2}{16\eta }%
\right) ^{-1}\left[ \frac{2r\lambda }{0.18\cdot 8\pi }\frac{\xi _0}L\left[
2\ln \left( 1-e^{-\frac L{\xi _0}\sqrt{\frac{3\lambda ^2}{16\eta }\frac R{%
0.18}}}\right) -2\sum_{n_1,n_2=1}^\infty \frac{e^{-L\sqrt{\frac{3\lambda ^2}{%
16\eta }\frac R{0.18}}\sqrt{n_1^2+n_2^2}}}{L\sqrt{n_1^2+n_2^2}}\right]
\right. \right.   \nonumber \\
&&\left. \left. +\frac{4r^2\eta }{0.18^2\cdot 32\pi ^2}\left( \frac{\xi _0}L%
\right) ^2\left[ 2\ln \left( 1-e^{-\frac L{\xi _0}\sqrt{\frac{3\lambda ^2}{%
16\eta }\frac R{0.18}}}\right) -2\sum_{n_1,n_2=1}^\infty \frac{e^{-L\sqrt{%
\frac{3\lambda ^2}{16\eta }\frac R{0.18}}\sqrt{n_1^2+n_2^2}}}{L\sqrt{%
n_1^2+n_2^2}}\right] ^2\right] \right\} .  \label{Tcritical4}
\end{eqnarray}
\begin{figure}[h]
\includegraphics[{height=9.0cm,width=9cm,angle=360}]{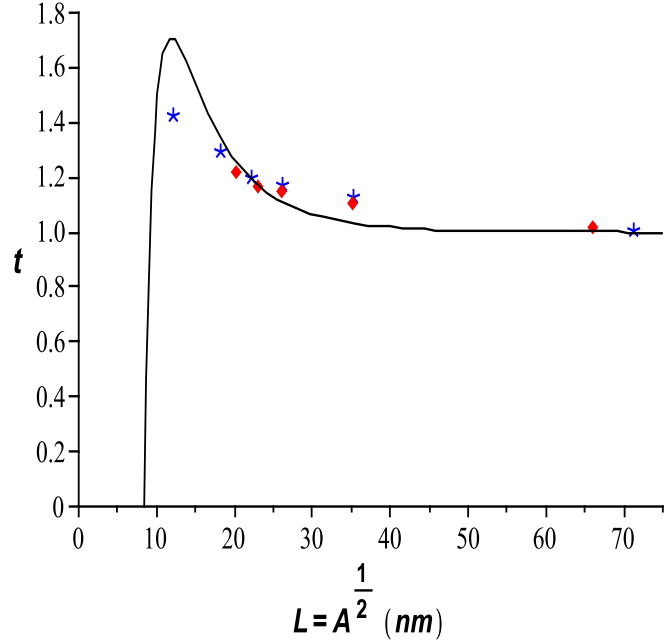}
\caption{ Reduced critical temperature $t= T_c^{\text{wire}}/T_{\text{bulk}}$ as a function
of the square root of the cross section $A^{1/2}$ (nm), from
  Eq.~(\ref{Tcritical4}) for an Al wire (the solid curve). The
diamond symbols are data from Ref.~\cite{shanenko} and the star
symbols are data from Ref.~\cite{zgirski}. We have used
$r=150\times 10^{4}$ and $R\approx 23$.}
\end{figure}

Notice that, due to the presence of the exponentials, the double series in
Eq. (\ref{Tcritical4}) is convergent. Therefore, they can be truncated at
some finite value for $n_1$ and $n_2$, so that a plot of the curve $T_c^{%
\text{wire}}(L)$ vs. $L$ can be drawn. In fact, the series are rapidly
convergent and no detectable difference exists if we take the sums over $n_1$
and $n_2$ up to 50 or a higher number. In Fig. 3 this curve is plotted and
compared with experimental data from Refs. \cite{shanenko,zgirski} for an
aluminum wire. Here, we have used the same tabulated values for $T_F$, $T_c$
and $v_F$ for aluminum as in the case of films. Also, for the parameter $R$
a larger value than the corresponding one for films ($R_{\text{wire}}=20\,R_{%
\text{film}}$) is taken. This is to account for the fact that samples in the
form of wires are more sensitive to the presence of impurities than in the
case of films \cite{tinkham}. From Fig. 3, we then notice that, for not
extremely thin wires, the data agrees quite well with the theoretical curve.
We see that the theoretical predicted behavior of the critical temperature
as a function of the square root of the cross section area (for us, the
transverse linear dimension $L$) is qualitatively of the same type we found
for films. Therefore, if one follows the same line of reasoning we have done
for films, one may conclude that the phase transition for these
superconducting aluminum wires is first order, just as for aluminum films.
This conjecture is reinforced, if one remembers that Eq. (\ref{seg}) for
second-order transitions is equally applicable to wires, showing a similar
behavior as that illustrated in Fig. 2, in which the curve approaches the
bulk critical temperature from below. However, it is clear from the data
that the critical temperature takes higher values as $L$ is decreased, thus
being incompatible with  the expected behaviour of a second order transition. 

\section{Concluding Remarks}

Studies on the dependence of the critical temperature for films with its thickness have been done in other contexts and approaches, different from  the one we adopt. For instance in Refs. \cite{cardy,zinn} an analysis of the renormalization group in finite-size geometries can be found.  Also, such a dependence has been investigated in \cite{quaterman,asanitsu,raffy,minhaj} from both experimental and theoretical points of view, explaining this effect in terms of proximity, localization and Coulomb interaction. In particular, Ref. \cite{quateman} predicts, as our model also do, a suppression of the superconducting transition for thicknesses below a minimal value. More recently in Ref. \cite{shanenko} the thickness dependence of the critical temperature  is explained in terms of a shape-dependent superconducting resonance, but no suppression of the transition is predicted or exhibited. 
 
In this paper we have adopted a phenomenological approach, discussing the $\left( \lambda |\varphi |^4+\eta
|\varphi |^6\right) _D$ theory compactified in $d\leq D$ Euclidean
dimensions.~We have presented a general formalism which, in the framework of
the Ginzburg--Landau model, is able to describe phase transitions for
systems defined in spaces of arbitrary dimension, some of them being
compactified. We have focused on the situations with $D=3$ and $d=1,2,3$,
corresponding (in the context of condensed-matter systems) to films, wires
and grains, respectively, undergoing phase transitions which are supposed to
be described by (mean-field) Ginzburg--Landau models. We have parametrized
the bare mass term in the form $m_0^2=\alpha (T/T_0-1)$, with $\alpha >0$
and $T_0$ being a parameter with the dimension of temperature, thus placing
the analysis within the Ginzburg--Landau framework. This generalizes
previous works dealing with first- and second-order transitions and
low-dimensional compactified subspaces \cite{linhares,malbouisson3,luciano3}%
. Such a generalization is far from being trivial, since it involves
extensions to several dimensions of the one-dimensional mode-sum
regularization described in Ref.~\cite{elizalde}. These extensions require,
in particular, the definition of symmetrized multidimensional
Epstein--Hurwitz functions with no analog in the one-dimensional case. It is
this kind of mathematical framework that allows us to obtain the general
formula (\ref{crittemp}), which may be particularized to films, wires and
grains, thereby implying the peculiar forms of the critical temperature as a
function of the linear dimension $L$, for the three physically interesting
cases.

It should be observed the very different form of Eqs.~(\ref{tcfilm}), (\ref
{tcwire}) and (\ref{tcgrain}) when compared with the corresponding ones for
second-order transitions given by Eq.~(\ref{seg}), obtained within the
Ginzburg--Landau $\varphi ^4$ theory. In all cases, the functional form 
of the dependence of the critical temperature $T_c(L)$ on the linear dimension
$L$ is of the following type: it grows from zero at a nonnull
minimal allowed value of $L$ below the bulk transition temperature $T_c$
as $L$ is enlarged, reaching a maximum above $T_c$ and afterwards starting to
decrease, going asymptotically to $T_c$ as $L\rightarrow \infty $. Eq.~(\ref{tcfilm}) is
in qualitatively good agreement with measurements \cite{strongin} taken for
a superconducting aluminum film, especially for thin ones. Moreover, experimental data
published in very recent years for an Al superconducting wire \cite{shanenko,zgirski}
show good accordance with Eq.~(\ref{tcwire}). Due to the extreme difficulties in preparing very thin wires,
however, there is an unfortunate lack of data for $L\lesssim 15$ nm, which
prevents the testing of the characteristic behavior of $T_c$ we expect in
this range of values, with a sudden drop to zero of $T_c$ after it reaches a
maximum value above the bulk one.
This is a very
contrasting behavior with that of the critical temperature for materials
displaying a second-order phase transition \cite{luciano2}, for which the
critical temperature increases monotonically from zero, again corresponding
to a finite minimal film thickness, going to the bulk transition temperature
as $L\rightarrow \infty $. Such behavior may indicate that from the form of
the dependence of the critical temperature on the size of the system, the
order of the transition the system undergoes could be inferred. 
Finnaly we should mention the important point that
{\it hysteresis} is a characteristic feature of a first-order transition. In our
case, it would mean that the transition temperature in the direction from
the normal phase to the superconducting phase (let's call it $T_c^{NS}$) is
different from the critical temperature in the inverse way ($T_c^{SN}$). To
our knowledge, the experiments investigating the thickness-dependent transition temperature are only from
the normal state to the superconducting one, so they do not show hysteresis.
We have not found, at least in the papers that are useful for us,
experimental studies on the comparison of the critical temperatures $T_c^{NS}
$ and $T_c^{SN}$. For us, remembering Eqs.(\ref{crittemp}) and (\ref{tc}), such a study would require separate calculations of the the Ginzburg-Landau phenomenological parameters $\lambda $, $\alpha $ and $\eta $ for the transition in the directions $NS$ and $SN$.

\acknowledgments This work has received partial financial support from CNPq
and Pronex. YWM is suported by FAPESP, grant 06/56653-9. APCM received partial support from FAPERJ. We thank E. Curado for  interesting discussions.

\end{document}